\documentclass[12pt]{article}

\usepackage{graphicx}
\usepackage{epsfig}

\newcommand{\rf}[1]{(\ref{#1})}
\newcommand{\bea}{\begin{eqnarray}}
\newcommand{\eea}{\end{eqnarray}}

\newcommand{\g}{\gamma}

\renewcommand{\l}{\lambda}
\renewcommand{\L}{\Lambda}
\renewcommand{\b}{\beta}

\newcommand{\n}{\nu}

\renewcommand{\th}{\theta}
\newcommand{\del}{\delta}

\newcommand{\oh}{\frac{1}{2}}

\newcommand{\ra}{\right\rangle}
\newcommand{\la}{\left\langle}
\newcommand{\prt}{\partial}
\newcommand{\mi}{\!-\!}

\newcommand{\cD}{{\cal D}}

\newcommand{\cO}{{\cal O}}
\newcommand{\cG}{{\cal G}}

\newcommand{\tZ}{{\tilde{Z}}}

\newcommand{\noi}{\noindent}

\def\void{}
\def\labelmark{}

\newenvironment{formula}[1]{\def\labelname{#1}
\ifx\void\labelname\def\junk{\begin{displaymath}}
\else\def\junk{\begin{equation}\label{\labelname}}\fi\junk}%
{\ifx\void\labelname\def\junk{\end{displaymath}}
\else\def\junk{\end{equation}}\fi\junk\labelmark\def\labelname{}}

{\ifx\void\labelname\def\junk{\end{array}\end{displaymath}}
\else\def\junk{\end{array}\right.\end{equation}}
\fi\junk\labelmark\def\labelname{}\def\junk{}
}

\newcommand{\beq}{\begin{formula}}
\newcommand{\eeq}{\end{formula}}
\newcommand{\beqv}{\begin{formula}{}}

\newlength{\footindent}
\makeatletter
\renewcommand{\@makefntext}[1]{\noindent\makebox[\footindent][r]{\@makefnmark}#1}
\makeatother
\begin{document}

\begin{flushright}
NBI-HE-98-23\\
MPS-RR 1998-33\\
BI-TP 98/39
\end{flushright}

\begin{center}
\vspace{24pt}
{ \large \bf Connected Correlators in Quantum Gravity}

\vspace{24pt}

{\sl J. Ambj\o rn}$\,^{a,}$\footnote{email ambjorn@nbi.dk},
{\sl  P. Bialas}$\,^{b,}$\footnote{email pbialas@physik.uni-bielefeld.de\\
permanent address: Institute of Comp. Science, Jagellonian University, 
30-072 Krakow, Poland}
and
{\sl J. Jurkiewicz}$\,^{a,}$\footnote{email jurkiewicz@nbi.dk, \\
permanent address: Institute of Physics, Jagellonian University, 30-059 Krakow,
Poland} 
\vspace{24pt}

$^a$~The Niels Bohr Institute, \\
Blegdamsvej 17, DK-2100 Copenhagen \O , Denmark, 

\vspace{24pt}
$^b$~Fakult\"{a}t f\"{u}r Physik, Universit\"{a}t Bielefeld, \\
33501 Bielefeld, Germany

\vspace{36pt}

\end{center}

%\addtolength{\baselineskip}{0.20\baselineskip}
\vfill

\begin{center}
{\bf Abstract}
\end{center}

\vspace{12pt}
\noi
We discuss the concept of connected, reparameterization 
invariant  matter correlators in quantum gravity. We
analyze  the effect of  discretization in two solvable cases~:
branched polymers and two-dimensional simplicial gravity. In both cases
the naively defined connected correlators for a fixed volume
display an anomalous behavior, which could be interpreted as a long-range
order. We suggest that this is in fact only a highly 
non-trivial finite-size effect and  propose an improved 
definition of the connected correlator, which reduces the effect. 
Using this definition we illustrate the
appearance of a long-range spin order in the Ising model on a two-dimensional
random lattice in an external magnetic field $H$, when $H \to 0$ and
$\beta=\beta_C$.

\vfill

\newpage

\section{Introduction}

In a theory where gravity is quantized it is non-trivial to define the
concept of a connected correlation function.  The problem is only
apparent once one has a genuine non-perturbative definition of quantum
gravity where one can go beyond the expansion around flat space. To
exemplify the problem let us define a reparameterization invariant
correlation function in quantum gravity: \beq{*1} G^{\phi_1\phi_2}_\L
(R) = \int \cD [g] \int\cD [X]\;e^{-S_\L}\; \int\int dV_g(\xi)\,
dV_g(\xi') \; \phi_1(\xi)\phi_2(\xi') \; \del(D_g(\xi,\xi')-R).  \eeq
Here $dV_g$ is the invariant volume element and $D_g(\xi,\xi')$ the
geodesic distance between $\xi$ and $\xi'$ for a given metric
$g$. $\cD [X]$ symbolizes the integration over additional degrees of
freedom of the theory, i.e.~ the matter fields and $\phi_i$ are local
observables built from these fields and/or the gravitational field.
The functional integral is over equivalence classes of metrics, i.e.\
the group of diffeomorphisms is divided out. Finally $\L$ is the
cosmological constant, i.e.\ we have the following decomposition of
the action: \beq{*1a} S_\L = \L V_g + S_V[g,X],~~~~~V_g = \int
dV_g(\xi), \eeq where $S_V$ is independent of the cosmological
constant.  As we can see from \rf{*1} the correlator involves three
rather than two non-trivial operators, contrary to the flat space
case, where the definition of a distance does not involve an extra
operator.  The new element is the geometric ``separator'', which for
any non-trivial geometry becomes a complicated non-local object.  A
partition function for quantum gravity can be defined by \beq{*2} Z_\L
= \int \cD [g] \int \cD [X] \;e^{-S_\L}, \eeq and
$G^{\phi_1\phi_2}_\L(R)$ is an {\it unnormalized} correlation
function.  The simplest object of this kind is a ``geometric''
two-point function, including only the ``separator'': \beq{*3}
G^{11}_\L (R) = \int \cD [g] \int \cD [X] \;e^{-S_\L}\; \int\int
dV_g(\xi) dV_g(\xi') \; \del(D_g(\xi,\xi')-R), \eeq i.e.\ the same
object as in \rf{*1}, just with $\phi(\xi) \to 1$.  $G^{11}_\L(R)$ can
be viewed as the partition function for the ensemble of universes
where two marked points are separated by a geodesic distance $R$. We
have \beq{*3.1} \int_0^\infty dR \; G^{11}_\L(R) = \frac{\prt^2}{\prt
\L^2} Z_\L \equiv \tZ_\L \eeq This two-point function plays a special
role in describing the geometric properties of the system
\cite{aw}. In general $G^{11}_\L(R)$ is expected to fall off
exponentially, reflecting the fact that there is an exponentially
small probability to create a universe where two points are separated
by a geodesic distance $R$ much larger than some power of the
cosmological constant. This power sets the geometric scale of the
system\footnote{The definitions \rf{*1}-\rf{*3.1} apply to any field
theory of quantum gravity. However, when we say that the geometric
scale is set by the cosmological constant, we have in mind
two-dimensional quantum gravity. In four-dimensional quantum gravity
the geometric structure might be more complicated since the
gravitational coupling constant is expected to set the scale of
quantum fluctuations and the cosmological coupling constant the scale
of the universe.}, or a relation between the average radius $\la R
\ra$ and the average volume $\la V\ra$ of the universe: $\la V\ra
\propto \la R\ra^{d_h}$, where $d_h$ is the Hausdorff dimension.  The
concept of a geometric scale $\la R \ra$ is very important when we
discuss the problem of matter correlations, which in principle may
involve some other ``physical'' scale.  The geometric scale is
controlled by the cosmological constant, while the physical scale will
in general depend on some other coupling constants, describing the
matter sector of the theory. Away from the critical point in this
sector we expect the physical scale to be small compared with the
geometric scale. Only then we can expect to distinguish those two
scales In the following we shall always assume that it is the case.
With a small abuse of notation, we call this limit {\it the
thermodynamic limit}.  Close to the critical point the two scales
become comparable and there we may expect to find a non-trivial
coupling between the matter and gravitational sectors of the
theory. If we view $G^{11}_\L(R)$ as a partition function, the
distance $R$ plays the role of an additional coupling constant of the
theory.  All these peculiarities make the definition of {\it
connected} correlators difficult and sometimes ambiguous.
%We must be very careful to define properly the
%range of $R$, which is physically important because extending $R$ to extreme
%values, while keeping the cosmological constant fixed will in general
%result in deforming the geometric properties of the system, which
%eventually will become a long ``one-dimensional'' tube. Clearly this is
%not a ``physical'' situation, by which we mean a system with a homogeneous
%distribution of volume. We can even visualize  ``phase transitions''
%when increasing $R$ changes the Hausdorff dimension $d_h$.
%To stay in the ``physical'' phase, even when 
%discussing the ``large $R$'' limit we should keep in mind that
%we must not exceed the geometric scale, $<R> < R << \infty$ or
%$R/<R>$ can be large, but remains to be  $\cO (1)$.

{\it A priori} we have 
various possibilities of defining a normalized correlation function.
Let us mention two different definitions:
\beq{*3.2}
\la \phi_1 \phi_2\ra_\L(R) = \frac{ G^{\phi_1\phi_2}_\L(R)}{\tZ_\L}
\eeq
and 
\beq{*4}
\la \phi_1 \phi_2\ra_\L(R) = \frac{G^{\phi_1\phi_2}_\L(R)}{G^{11}_\L(R)}.
\eeq
The definitions differ by the way geometry is counted. In \rf{*4} it is 
counted in the same way in numerator and denominator. 
In fact, if $\phi_1$ and $\phi_2$ are independent observables which do 
not couple to gravity at all, definition \rf{*3.2} gives
\beq{*4.1}
\la \phi_1 \phi_2\ra_\L(R) = \la \phi_1 \ra \la \phi_2\ra 
\frac{G^{11}_\L(R)}{\tZ_\L},
\eeq 
 On the other hand,
the definition \rf{*4} leads to
\beq{*4.2}
\la \phi_1 \phi_2\ra_\L(R) = \la \phi_1 \ra \la \phi_2\ra,
\eeq
independent of $R$.

The results \rf{*4.1} or \rf{*4.2} are trivial from the 
point of view of the  correlation functions.
Both should be viewed as examples of a situation, where the observables
$\phi_i$ at different points are uncorrelated. The correlator can in this
case be expressed as a product of three single-operator averages,
which is best seen in \rf{*4.1}. 
One would expect a similar structure of the correlation function
in the {\it thermodynamic limit}
when the observables $\phi_i$ do couple to gravity,
but when the distance between the two points $R$ is 
much bigger than the correlation length. The expectation values
$\la \phi_i\ra$ should then be replaced by $\la \phi_i \ra_\L$ implying
the non-trivial dependence on the cosmological constant. 
The $R$ dependence should factorize, like in \rf{*4.1} or \rf{*4.2}.

At smaller distances there may be deviations from a simple factorization
\footnote{An attempt to derive a systematic operator product expansion can be 
found in reference \cite{operator}.}.
The obvious questions are: 
Can these be interpreted as a signal of a correlation between the fields?
Which type of behavior can still be attributed to the uncorrelated
operators? Such questions clearly can not be answered without some
hint from a solvable model. Unfortunately the number of solvable theories,
where one can actually compute the two-point functions,
is very limited. In the next section 
we shall present some exact results concerning operators
which we believe are uncorrelated.

Of course a behavior like \rf{*4.1} and \rf{*4.2}
can not be observed
if we have correlators with infinite correlation 
length (or where the geometric and physical scales become comparable), 
as will be the case when we consider correlators between the conformal fields
coupled to two-dimensional quantum gravity. 

To obtain a consistent definition of the matter correlation length we
must  define the concept of a {\it connected}
correlation function. Various definitions have been 
given in the literature \cite{jansmit,bielefeld,bialaspro}
and below we summarize the discussion. First we face the problem of a 
sensible definition of $\la \phi \ra_\L$. Several choices seem possible,
but in order to match the definitions of correlation functions, 
as given by \rf{*3.2} or  \rf{*4} one can use either
\beq{*4.3}
\la \phi \ra_\L = \frac{1}{\tZ_\L} 
\int \cD [g] \int \cD [X]\;e^{-S_\L}\; {V_g}\int dV_g(\xi) \, \phi(\xi)=
\frac{\int_0^\infty dR\; G^{\phi 1}_\L(R)}{
\int_0^\infty dR\; G^{1 1}_\L(R)}
\eeq
or 
\beq{*4.4}
\la \phi \ra_\L(R)\equiv \la \phi \, 1 \ra_\L (R) 
= \frac{G^{\phi \, 1}_\L(R)}{G^{1 1}_\L(R)}
\eeq
The last definition depends
on the geodesic distance $R$, and it is from this point of 
view slightly unusual as a definition of an expectation 
value of a field. But not much more than in usual field 
theory where the lack of translational invariance caused by 
an external field can introduce a space-time dependence in
$\la \phi \ra$. As discussed above, such dependence is expected
to disappear for large $R$ implying the existence of
a well defined limit $\la \phi \ra_{\L}(R \to \infty)$. 
In fact, this limit should agree with the value defined
by \rf{*4.3} provided the correlation length of the 
matter fields is (much) less than the average radius of 
the universe. 
The definition
of $\la \phi \ra_{\L}$ is related to the correlation between 
the $\phi$ and unit
operators, as is clear from \rf{*4.3} and \rf{*4.4}. 
Thus the statement that the two definitions of $\la \phi \ra$ agree
in the thermodynamic limit, for large $R$ 
is equivalent to
the statement that we have a factorization
\beq{*zfac}
G^{\phi\, 1}_\L (R) \approx \la \phi \ra_\L G^{1 1}_\L(R)
\eeq
for such  values of $R$. In general we would expect the unit operator
to be uncorrelated with any other operator, so naively we
would expect \rf{*zfac} to be satisfied almost for any value of $R$.
This can be checked numerically. 

Note that the correlation function $G^{\phi 1}_\L(R)$
almost inevitably enters in any sensible definition of the 
connected part of a correlator since one has to consider 
an object like
\beq{*4.3a}
\la \phi_1 \phi_2\ra^{con}_\L (R) \equiv 
\la (\phi_1 - \la \phi_1\ra 1)(\phi_2 - \la\phi_2\ra 1)\ra_\L (R),
\eeq
with some definition of $\la \phi_i \ra$. If we use \rf{*4.4} as a definition
of $\la \phi_i \ra$ and \rf{*4} as a definition of the correlator, then 
eq.\ \rf{*4.3a} fulfills the standard decomposition and can be written as
\beq{*4.4a}
\la \phi_1 \phi_2\ra_\L(R) - \la\phi\ra_\L(R)\la\phi_2\ra_L(R).
\eeq
If we use \rf{*3.2} and \rf{*4.3} instead, eq.\ \rf{*4.3a} can be 
written as 
\bea\label{*4.5a}
\lefteqn{\la \phi_1 \phi_2\ra^{con}_\L (R) =}\\ \nonumber 
&&\frac{1}{\tZ_\L} \; (G^{\phi_1\phi_2}_\L(R)-
\la \phi_1\ra_\L G^{1\phi_2}_\L(R)- \la \phi_2\ra_\L G^{1\phi_1}_\L(R)   + 
\la \phi_1 \ra_\L\la\phi_2\ra_\L G^{11}_\L(R).
\eea
In this case one does not have the standard local decomposition as in 
eq.\ \rf{*4.4a}, but integrating over $R$ one obtains:
\beq{*4.6a}
\la \phi_1 \phi_2\ra_\L - \la \phi_1 \ra_\L\la\phi_2\ra_\L,
\eeq
where the first term in \rf{*4.6a} is the usual integrated 
correlator as known from two-dimensional quantum gravity:
\beq{*4.7a}
\la \phi \phi \ra_\L = \frac{1}{\tZ_\L}\int \cD [g]\int \cD [X] \; e^{-S_\L} \;
\int \int dV_g(\xi) dV_g(\xi') \; \phi(\xi)\phi(\xi').
\eeq
We can now define {\it the correlation length} by  the exponential
decay of  $\la \phi \phi\ra_\L^{con}(R)$, defined either by \rf{*4.4a}
or \rf{*4.5a} and the thermodynamic limit is where this 
correlation length is much smaller than the average radius of the 
the universe.

The  fixed volume partition functions 
$\tZ_V$ and $G^{\phi_1\phi_2}_V(R)$ are 
related to $\tZ_\L$ and $G^{\phi_1\phi_2}_\L(R)$ by Laplace transformations
\beq{*5}
\tZ_\L = \int^{\infty}_0 dV \;e^{-\L V} Z_V,~~~~~~
G^{\phi_1\phi_2}_\L(R) = \int^{\infty}_0 dV\; G^{\phi_1\phi_2}_V(R).
\eeq
and it is natural to use the definition of correlators, expectation
values of fields and connected correlators corresponding to  eqs.\ 
\rf{*3.2}, \rf{*4.1}, \rf{*4.3} and \rf{*4.5a}, just with the
the partition functions for fixed $\L$ replaced by the ones for fixed 
volume $V$, given by \rf{*5}. Note in particular that we have
\beq{*10}
\la \phi \ra_V = \frac{1}{ Z_V}\int \cD[g]\,\del(\int dV_g(\xi)\mi V) 
\; \int\cD [X] \;\;e^{-S_V[g,X]}\;\frac{1}{V_g} \int dV_g(\xi) \;\phi(X;\xi),
\eeq 
as one would have expected. The connected correlator for 
finite volume could thus be defined as 
\bea\label{*11}
\lefteqn{\la \phi_1 \phi_2\ra^{con}_V (R)  =} \\ \nonumber
&& \frac{1}{V^2Z_V} \; \Big(G^{\phi_1\phi_2}_V(R) - 
\la \phi_1\ra_V G^{1\phi_2}_V(R)-\la \phi_2\ra_V G^{1\phi_1}_V(R)+
\la \phi_1 \ra_V \la \phi_2\ra_V  G^{11}_V(R)\Big),
\eea
with the normalized fixed volume correlators defined as
\beq{*11a}
\bar{G}^{11}_V(R)=\frac{1}{V^2Z_V}G^{11}_V(R),
\eeq
satisfying
\beq{*11b}
\int_0^{\infty}dR\; \bar{G}^{11}_V(R) = V.
\eeq

As we show in the next Section, the discrete regularization of a theory may,
and in fact does lead to some complications, where the finite volume
effects tend to mimic the physical correlations even in situations where
there are no correlations in the grand canonical formulation. This was first
realized in \cite{bialaspro}.

% One expects
%$\la\phi_1\phi_2\ra_V^{con}(R)$ to behave as 
%\beq{*16}
%\la\phi_1\phi_2\ra_V^{con}(R) =V^{1-1/d_h-\Del_{\phi_1}-\Del_{\phi_2}} F(x),
%~~~x= \frac{R}{V^{1/d_h}},
%\eeq 
%where $\Del_{\phi_{1,2}}$ are scaling dimensions of the fields $\phi_{1,2}$
%{\it after} coupling to quantum gravity, where $x$ is dimensionless (thus
%defining the possibility of an 
%anomalous scaling of the geodesic distance $R$ relatively to
%to the volume of space-time, parameterized by the intrinsic Hausdorff
%dimension $d_h$). While these definitions are obviously modeled after 
%two-dimensional quantum gravity, they apply in principle to any definition of 
%higher dimensional Euclidean gravity, 
 
\section{Analytical results}

There are unfortunately very few systems, where the concepts presented
above can be compared with the {\em analytic} prediction. We present 
here the few models 
where the two-point function can be explicitly calculated. These systems
are branched polymers \cite{adfo,bbj} and 
two-dimensional simplicial gravity \cite{aw,ajw,jj}. 
In both cases one defines a discretized (integer)
geodesic distance $r$ between the points of the manifold. The physical
relation between the the continuum and discrete geodesic distances is
\beq{*14}
R = r a,
\eeq
with the lattice spacing $a \to 0$ in the continuum limit, but the
distance $R$ kept fixed. In both cases the relation between the
the continuum volume $V$ and the discrete volume $N$ 
is anomalous \footnote{One usually chooses the scaling $R = a^{2/d_h} r$ 
and $V= a^2N$ rather than \rf{*14} and \rf{*13a}. However, 
\rf{*14} is more  convenient from a notational point of view in the 
arguments to follow, so we prefer to work with a ``rescaled'' cut-off 
$a$ defined by \rf{*14} and \rf{*13a}.}:
\beq{*13a}
V = N a^{d_h},
\eeq
where $d_h$ is the Hausdorff dimension of the system, equal 2 for a
generic branched
polymer and 4 for two-dimensional simplicial gravity.

In both cases there is no extra matter content in the theory and the
only observables we can discuss are related to the local geometric
properties of the manifold. In the case of simplicial quantum gravity
these can be some functions of the coordination number of a vertex, in
the case of branched polymers -- functions of the branching ratio at a vertex.
Both cases correspond to the observables, which we expect to be
essentially uncorrelated for $r > 0$.

\subsection{Branched polymers}
Let us start by repeating  
the discussion of the simpler case of the branched polymers
\cite{adfo,bbj}. The
partition function in this model is given as a weighted sum over the
ensemble of trees.  Trees are weighted by one-vertex branching
weights $p(q_i)$. The partition function for the ensemble of trees
is given by~:
\beq{zn} 
\bar{Z}=\sum_N \exp(-\mu N)\;\sum_{T\in {\mathcal T}_N} \frac{1}{C(T)}
\prod_{i\in T} p(q_i)
\eeq
where $q_i$ is the order of a vertex $i$, $N$ is the number of vertices 
and $C(T)$ is an appropriate symmetry factor of the graph; $\mu$ 
plays the role of the bare cosmological constant.

The correlation functions are constructed by means 
of the partition function $Z$
of {\em planted, rooted, planar} trees. This partition
function can be found from the following recursive
equation~\cite{adfo}: 
\begin{equation}\label{zgen}
Z=e^{-\mu}F(Z)
\end{equation}

where 
\beq{*defbp}
F(Z)=\sum_{q=1} p(q) Z^{q-1} \, .
\eeq
In the generic case equation \rf{zgen} has a critical point $\mu_C$
for which 
\beq{}
e^{-\mu_C}F'(Z_0) = 1,~~~~~Z_0=Z(\mu_C).
\eeq
For $\mu$ approaching the critical value $\mu_C$ 
from above, $Z$ has the following singularity~:
\beq{zexp}
Z(\mu)=Z_0-Z'_1\sqrt{\mu-\mu_c} + \cO (\mu-\mu_c)
\eeq

The natural definition of a distance $r$ between two vertices
of a graph is the number of links joining these vertices.
The discrete analogue of the geometric two-point function
can be calculated in terms of $Z(\mu)$ \cite{bialaspro,adfo,bbj} as
\beq{*bp}
G_{\mu}^{11}(r)
=\Big(1+\frac{Z}{\partial_\mu Z}\Big)^{r-1}Z^2.
\eeq
The factor $Z^2$ in \rf{*bp} is a contribution from the two ends of the chain.
Close to the critical point
\beq{*DELTA}
\Delta = -\log(1+\frac{Z}{\partial_\mu Z})\propto \sqrt{\mu-\mu_c}
+\cO(\mu-\mu_c).
\eeq
and \rf{*bp} becomes
\beq{*bp1}
G_{\mu}^{11}(r)
=\exp(-(r-1)\Delta)Z(\Delta)^2.
\eeq

In a branched polymer system the only local observables we can construct are
functions of  vertex orders $q_1, q_2$ at the end points.
Replacing  $Z \to Z~e^{\l_i} F(Z e^{\l_i})/F(Z)$ at the end points
we obtain a generating function of all such observables: differentiating with
respect to $\l_i$  at $\l_i=0$  we generate
powers of $q_i$, $i=1,2$. Notice that for $Z$ satisfying \rf{zgen}
the generating function
\beq{aver}
\cG(\l)=\frac{e^\l F(Z e^\l)}{F(Z)} = \sum_n \la q^n\ra_{\mu} \frac{\l^n}{n!}
=\la \hat{\cG}(\l)\ra_{\mu},
\eeq
where the averages are taken with respect to the partition function $Z(\mu)$.
We conclude that for every choice of observables we have a simple
factorization 
\beq{*fact}
G_{\mu}^{\{\l_1,\l_2\}}(r) = \cG(\l_1)\cG(\l_2) G_{\mu}^{11}(r),
\eeq
which proves a lack of correlation between any pair of vertex order
operators at points $1$ and $2$. 

Near the critical point the generating function $\cG(\l)$ can be
expanded as
\beq{*expg}
\cG(\l)= \cG_0(\l)+\Delta \cG_1(\l) + \cO(\Delta^2)
\eeq
Using the simple form \rf{*bp1} of $G_{\mu}^{11}(r)$ we get
\bea
G_{\mu}^{\{\l_1,\l_2\}}(r)&=&
\Big(\cG_0(\l_1)+\cG_1(\l_1)\partial_r+\cdots\Big)
\Big(\cG_0(\l_2)+\cG_1(\l_2)\partial_r+\cdots\Big)G_{\mu}^{11}(r) \nonumber\\
&\approx &
\cG_0(\l_1)\cG_0(\l_2)G_{\mu}^{11}(r+\delta(\l_1)+\delta(\l_2)),
\label{*fact1}
\eea
where we have introduced a {\it shift} 
\beq{*fact3a}
\delta(\l)=\frac{\cG_1(\l)}{\cG_0(\l)},
\eeq
and where the dots stand for higher order terms typically proportional to
$d^2G_{\mu}^{11}(r)/dr^2$.

\subsection{2d gravity}

Two-dimensional simplicial gravity can be obtained as the planar limit of the
$\phi^3$ matrix theory. The coupling constant $g = e^{-\mu}$ of this
theory can be parametrized as
\beq{*x1}
8g^2 = s(1-s^2).
\eeq
In this parameterization $g=0$ corresponds to $s=1$ and the critical value
$g^2_c = 1/12\sqrt{3}$. In the planar limit each $\phi^3$ vertex is dual
to a triangle and a $\phi^3$ graph can be viewed as a two-dimensional surface
built from triangles, expansion in powers of $g$ becoming the expansion in
the area of the surface.
Following \cite{jj} we construct the transfer matrix \cite{transfer}
\beq{*x2}
G_\mu(X,Y;r)=\sum_{L,L'}X^L Y^{L'}G_\mu(L,L';r),
\eeq
where $G_\mu(L,L';r)$ is the sum of all possible connected
planar $\phi^3$ graphs with boundaries being the 
loops (planarly ordered   sequences of external links)
with lengths $L$ and $L'$, separated by a ``distance'' $r$. The
distance between the two vertices of a graph is the length of the
shortest path following the links of a graph. 
The transfer matrix $G_\mu(X,Y;r)$ can be calculated using the ``peeling''
method \cite{watabiki} giving
\beq{*transfer}
G_\mu(X,Y;r)=\frac{f_{\mu}(\hat{X})}{f_{\mu}(X)}\frac{1}{1-\hat{X}Y},
\eeq
where
\beq{fmu}
f_{\mu}(X,g)=(1-\frac{s X}{g})\sqrt{1-\frac{4g X}{s^2}}
\eeq
and $\hat{X}(X,r)$ satisfies
\bea\label{char}
r&=&-\frac{1}{\Delta}\log \frac{T(\hat{X})}{T(X)},\\ \nonumber
\Delta&=&s\sqrt{1-\frac{4g^2}{s^3}}.\\ \nonumber
T(X)&=&\frac{1-\frac{\Delta}{\sqrt{s^2-4gX}}}
{1+\frac{\Delta}{\sqrt{s^2-4gX}}}.
\eea
For $g \to g_c=e^{-\mu_c}$ we have 
\beq{}
\mu-\mu_c \propto \Delta^4 + \cdots
\eeq
which reflects the fact that the Hausdorff dimension $d_h=4$.
Introducing $\hat{T}=T(\hat{X})$ and $T(X)=e^{-\Delta r_0(X)}$ we have
\beq{*char}
\hat{T}=e^{-\Delta(r+r_0(X))}.
\eeq
Using this parameterization we easily express $f_{\mu}(\hat{X},g)$ as
\bea\label{fhx}
f_{\mu}(\hat{X},g)&=&
\frac{\Delta^3}{g^2}\coth({1\over 2}\Delta(r+r_0))
(1-\coth^2({1\over 2}\Delta(r+r_0)))\\ \nonumber
&=& F(\Delta(r+r_0)).
\eea
The transfer matrix described above can be used to calculate the
two-point correlators \cite{aw,operator}. 
Below we shall discuss only the simpler case,
when one of the operators is the unit operator. The solution 
for the general case with two non-trivial operators can be deduced from
symmetry. The first step will
be to close the incoming loop. We are still left with the open
final loop. The resulting function will be denoted by $\bar{G}_{\mu}(X;r)$
and can be expressed as
\bea\label{*step1}
\bar{G}_{\mu}(X;r)
&=& \oint_{C_Y}\frac{dY}{2\pi{\rm i}Y}G_{\mu}(X,\frac{1}{Y};r)Y\frac{\partial}
{\partial Y}\Psi(Y)\\
&=&
\frac{f_{\mu}(\hat{X})}{f_{\mu}(X)}
\hat{X}\frac{\partial}{\partial\hat{X}}\Psi(\hat{X}),
\eea
where
\beq{solution}
\Psi(X)=\frac{1}{2}(\frac{X}{g}+1)+\frac{1}{2}(1-\frac{s X}{g})
\sqrt{1-\frac{4 g X}{s^2}}
\eeq
is a generating function of the connected Green's functions of the
$\phi^3$ matrix model \cite{zuber}.
For small $\Delta$ the effect of closing the incoming loop can be
represented as
\bea\label{*step2}
\bar{G}_{\mu}(X;r)&=&\frac{F(\Delta(r+r_0))-\partial_r F(\Delta(r+r_0))+\cdots}
{F(\Delta r_0 )}\\ \nonumber
&\approx& \frac{F(\Delta(r-1+r_0))}{F(\Delta r_0)}.
\eea
The dots correspond to  terms proportional to the second derivative of $F$.
The $X$ dependence in this formula appears through $r_0=r_0(X)$.  
This is exactly the information we need. The particular choice of a
set of operators we wish to study is not very important, provided the
incoming lines represented by the $X$ dependence are attached to
some local geometric object. The simplest choice is just a line, joining
the end points, which leads to the generating function
\beq{*D2}
G_{\mu}^{\{\l_1,\l_2=0\}}(r)= \bar{G}_\mu(X=g~e^{\l_1};r),
\eeq
where we introduced, as before, the parameter $\l_1$ to count the length
of the line and its higher moments. Notice that although the $r$ dependence
of this quantity is more complicated than in the branched polymer case,
the fundamental properties remain the same. Denoting  by 
$\delta(\l)=r_0(\l)-r_0(0)$ we have
\beq{*D2bis}
G_{\mu}^{\{\l_1,\l_2=0\}}(r)=\cG(\l_1) G_{\mu}^{11}(r+\delta(\l_1)).
\eeq
where
\beq{}
\cG(\l)=\frac{F(\Delta r_0(0))}{F(\Delta r_0(\l))}.
\eeq
Again the correlation function
involving the nontrivial operator ($\l \ne 0$) is related to that of the
unit operator ($\l=0$) by a $\l$-dependent multiplicative factor and a 
$\l$-dependent shift of $r$.

\subsection{The Ansatz for a connected correlator}
There are two important lessons one can learn from the 
examples presented above. 
The first one is that even in cases where there is no correlation the naive
factorization of the two-point correlator:
\beq{*naive}
G_{\mu}^{1 \phi}(r) = \la \phi\ra_{\mu} G_{\mu}^{11}(r)
\eeq
is satisfied only asymptotically, for large enough $r$. This does not
necessarily mean that there is no simple factorization, 
as we see from the example of pure 2d gravity, (see \rf{*char}). 
The object for which we observe a simple behavior is not the two-point
function itself, but rather a geometric object, natural to the evolution
of the characteristic equation of the transfer matrix. The two-point
function is itself expressible in terms of this object. Near the critical
point the geometric scale of the system is provided by $\Delta$ and the
relation between the cosmological constant and this parameter is anomalous
\beq{}
\mu-\mu_c \propto \Delta^{d_h}.
\eeq
In general only for $r$ large enough the correlation function 
$G_{\mu}^{11}(r)$ has a simple
exponential form.

The second lesson is that we can nevertheless extend the idea of factorization
to be satisfied for the whole range of $r$ if instead of \rf{*naive}   
we use the modified Ansatz:
\beq{*clever}
G_{\mu}^{1 \phi}(r)= (A_{\phi} - B_{\phi}~ \partial_r +\cdots )G_{\mu}^{11}(r)
\eeq
which asymptotically (for $r$ large enough for $G_{\mu}^{11}(r)$ to be purely
exponential) corresponds to
\beq{}
\la \phi\ra_{\mu}= A_{\phi} + B_{\phi}~\Delta + \cdots
\eeq
To  order $\cO(\Delta)$ the Ansatz \rf{*clever} can be viewed as an 
additional shift: 
\beq{*clever1}
G_{\mu}^{1 \phi}(r)= A_{\phi}~ G_{\mu}^{11}(r+\delta_{\phi})
\eeq
with
\beq{}
\delta_{\phi}=\frac{B_{\phi}}{A_{\phi}}.
\eeq
For two {\em uncorrelated} operators the corresponding Ansatz would be
\beq{*clever2}
G_{\mu}^{\phi_1 \phi_2}(r)= A_{\phi_1}~A_{\phi_2}~
G_{\mu}^{11}(r+\delta_1+\delta_2)
+\cO(\Delta^2).
\eeq
The additional shifts are specific for the operator and are additive. 
We also made an assumption that $A_{\phi_i} \ne 0$. 
The uncorrelated contribution
described above has to be subtracted if one is interested in the connected
correlation function. 
The effect described above is a finite-size correction. In the continuum
limit $r \to R/a$ and $\partial_r \to a~\partial_R$. It is however very
important if we analyze the system using the fixed volume {\em canonical}
numerical simulations.

The unnormalized correlation function $G_{\mu}^{\phi_1\phi_2}(r)$ can be
represented as a (discrete) Laplace transform of the fixed-$N$ contributions as
\beq{*sumc}
G_{\mu}^{\phi_1\phi_2}(r)=\sum_N \exp(-\mu N)\;C_N~G_{N}^{\phi_1\phi_2}(r),
\eeq
where $C_N$ controls the normalization of the fixed-volume 
correlation function and is independent of $\phi$. 
The commonly used normalization corresponds to
\beq{*norm}
\sum_r~G_{N}^{11}(r)=N.
\eeq
Using standard arguments $G_{N}^{\phi_1\phi_2}(r)$ can be obtained by inverting
the Laplace transform \rf{*sumc}. From \rf{*fact1} it follows that
\beq{*fact2}
G_{N}^{\phi_1\phi_2}(r)=\Big(A_{\phi_1}-B_{\phi_1}\partial_r+\cdots\Big)
\Big(A_{\phi_2}-B_{\phi_2}\partial_r+\cdots\Big)G_{N}^{11}(r)
\approx A_{\phi_1}~A_{\phi_2}~G_N^{11}(r+\delta_1+\delta_2).
\eeq
In the large-$N$ limit we expect
\beq{*scaling} 
G_{N}^{11}(r) \approx N^{1-1/d_h}g(x),
\eeq 
where the scaling variable $x=(r+\delta_0)/N^{1/d_h}$. 
It is clear from \rf{*fact2} that derivatives
$\partial_r$ produce finite-size corrections $\cO(1/N^{1/d_h})$ or
$\cO(1/L)$, where $L$ is the linear (discrete) size of the system.

In a numerical experiment at a fixed volume $N$ one measures the correlation
functions $G_N^{11}(r)$, $G_N^{1\phi_i}(r)$ and $G_N^{\phi_1 \phi_2}(r)$,
where $G_N^{11}(r)$ is normalized as in \rf{*norm}.
As we argued above we expect to $\cO(1/L^2)$
\beq{*phi-1}
G_N^{1\phi_i}(r)=A_{\phi_i} G_N^{11}(r+\delta_i).
\eeq
This behavior can easily be checked in numerical experiments and we
find it satisfied with amazing accuracy, not only in cases discussed above,
but also for other two-dimensional systems, as will be discussed in the next
section as well as in correlation experiments in four-dimensional simplicial
gravity \cite{future}. Following \rf{*phi-1} one would expect $A_i$ to be
an $N$-independent constant  $A_{\phi_i} = \la \phi_i \ra_{N=\infty}$.
Experimentally we found that an excellent fit to \rf{*phi-1} corresponds to
the choice $A_{\phi_i} = \la \phi_i \ra_N$, which for an extensive quantity 
differs by $\cO(1/N)$. 

In general  the corrrelation function  $G_N^{\phi_1 \phi_2}(r)$ may
contain the non-trivial connected part. However if we define the connected
part of the correlator in the analogy with \rf{*11} as
\bea\label{*11bis}
\lefteqn{G_N^{\phi_1\phi_2,conn}(r)  =} \\ \nonumber
&& G^{\phi_1 \phi_2}_N(r) - 
\la \phi_1\ra_N G^{1\phi_2}_N(r)-
\la\phi_2\ra_N G^{1\phi_1}_N(r)+ \\ \nonumber
&& \la \phi_1\ra_N \la\phi_2\ra_N  G^{11}_N(r),
\eea
even in case when there is no correlation in the grand canonical ensemble, as
discussed above, we find a non-zero contribution
\bea\label{*11again}
\lefteqn{G_N^{\phi_1\phi_2~conn}(r)=}\\ \nonumber
&& B_{\phi_1}~B_{\phi_2}~\partial_r^2 G_N^{11}(r)=B_1 B_2 N^{1-3/d_h} f(x),
\eea
where $f(x)=g^{''}(x),~x=(r+\delta_0)/N^{1/d_h}$. The measured correlation
mimics a real, physical scaling correlation function, corresponding to
an object with the anomalous dimension $2/d_h$.
Note that this behavior will be observed even if the two-point function
$G_{\mu}^{11}(r)$ is purely exponential, as is the case for the branched
polymer system. Following the discussion above it represents in fact
only the contribution from the {\em disconnected} part of the correlation 
function and can be viewed as an
artifact of the Laplace transform to a finite volume $N$. It is exactly this
part, which we would like to eliminate, when we try to measure the 
connected part of the correlation function. 

Our discussion shows that it may be impossible to eliminate completely
the disconnected contribution in a finite volume correlation experiment.
However, we may reduce it. The following simple redefinition of the
connected correlator: 
\bea\label{*11final}
\lefteqn{G_N^{\phi_1\phi_2,conn}(r)  =} \\ \nonumber
&& G^{\phi_1 \phi_2}_N(r) - 
\la \phi_1\ra_N G^{1\phi_2}_N(r+\delta_1)-
\la\phi_2\ra_N G^{1\phi_1}_N(r+\delta_2)+ \\ \nonumber
&& \la \phi_1\ra_N\la \phi_2\ra_N  G^{11}_N(r+\delta_1+\delta_2),
\eea
gives an extra factor $N^{-2/d_h}$ as compared with \rf{*11bis} and in
all practical cases was sufficient to reduce the disconnected signal below the
error level. This point will be discussed further in the next section.

\section{A numerical recipe}

In this section we present a simple algorithm, which permits us to
measure non-trivial correlators in numerical experiments. Assume that
from the experiment we know four correlation functions~: 
$G_N^{\phi_1\phi_2}(r)$, $G_N^{1~\phi_1}(r)$, $G_N^{1~\phi_2}(r)$ and 
$G_N^{11}(r)$.
To make use
of the formula \rf{*11final} we need four parameters: $\la\phi_1\ra_N$,
$\la\phi_2\ra_N$, $\delta_1$ and $\delta_2$. 

As a first step we use \rf{*phi-1}:
\beq{*phi-2}
G_N^{1\phi_i}(r)=\la\phi_i\ra_N G_N^{11}(r+\delta_i).
\eeq
Correlation functions measured in the experiment are given only for
integer values of the distance $r$. To interpolate between these values
we use a four-point interpolation formula, which is used to obtain a function
for $k > r > k+1$ from it's values at $k-1,~k,~k+1$ and $k+2$. Two
parameters $\la\phi_i\ra_N$ and $\delta_i$ are fitted by the least square
method. We decided to use this method rather than the $\chi^2$ since
including the numerical errors has an effect of giving a large importance
to the ends of the distribution, where both the correlation functions and
errors are small. In practice it turns out that the fitting is fairly
insensitive to the extrapolation method (we used the spline method as an
alternative). The value of $\la\phi_i\ra_N$ obtained in a fit was
practically the same as the average obtained independently from the experiment.
After some checks we decided to use the experimental value from the start
and to use \rf{*phi-2} only for a one-parameter fit 
to find the optimal value for $\delta_i$. One should
stress that in all cases 
the fit is very good provided $\la\phi_i\ra_N$ is not to small. From our
discussion it follows that the connected part of the correlation function
remains unchanged if we add a constant to the operator. We used this
property, when necessary to avoid  numerical difficulties.

\begin{figure}
\begin{center}
\includegraphics[width=11cm,,bbllx=51,bblly=60,bburx=516,bbury=434]{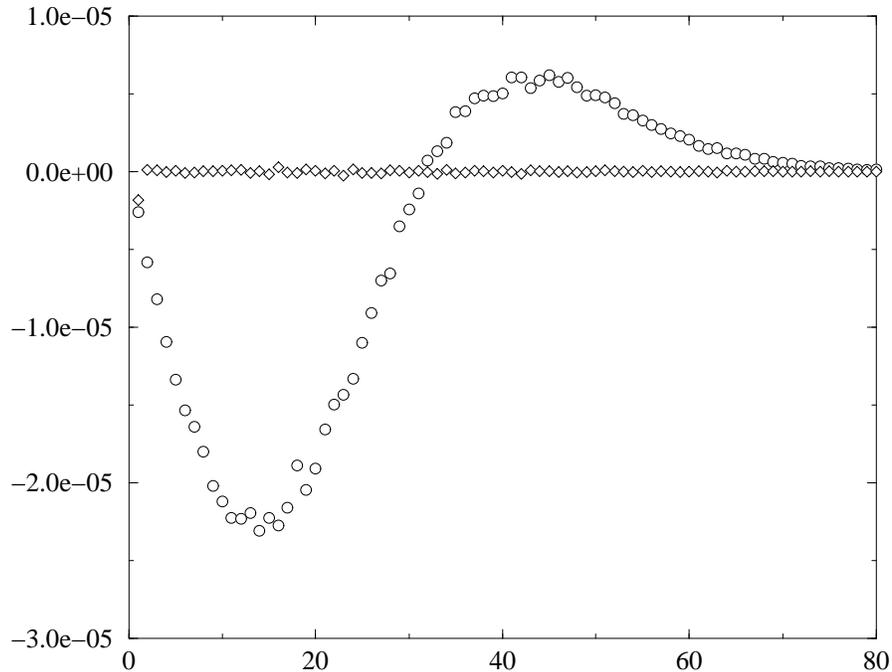}
\end{center}
\caption{\label{bp} Branched polymer.}
\end{figure}

We start our presentation
with the numerical results for the branched polymer model
for a system with 1024 vertices and weights $p(q)=1/q$. 
The results for the connected correlators between the branching ratios
at two points separated by a distance $r$ defined using \rf{*11bis} and 
\rf{*11final} are presented on the figure~\ref{bp}. 
This example illustrates the power of the proposed algorithm. We can see
that a strong disconnected signal is reduced almost completely and that
we are left with only local correlation at $r=0$.

\begin{figure}
\begin{center}
\includegraphics[width=11cm,bbllx=51,bblly=60,bburx=516,bbury=434]{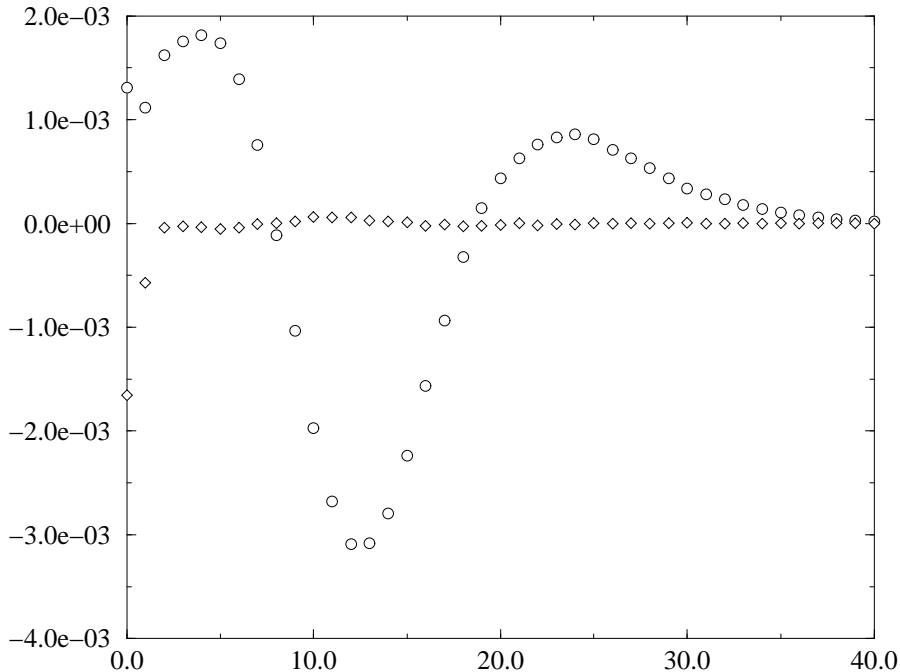}
\end{center}
\caption{\label{s2d} 2d gravity.}
\end{figure}

Similar effect is presented
on the  figure~\ref{s2d} for pure 2d gravity
using the data from the combinatorial spherical surfaces 
with regular triangulations
and with 8000 points. The
definition of a distance used in this measurement is different than
in the discussion presented in the section 2. Instead of the distance between
the $\phi^3$ vertices (centers of the triangles of the surface), defined
in terms of dual links, we use the distance between the vertices of the
surface, defined as a shortest path following the links of a surface.
We measure the connected correlator of the orders of two vertices (numbers
of triangles containing this vertex). 
This example shows that also in this case we observe the anomalous long-range
behavior if the 
the connected correlator is defined using \rf{*11bis}.
Using an improved definition
\rf{*11final} we observe a nontrivial correlation only for very small $r$. 

We used the same method to study the spin-spin correlations for the 
ferromagnetic Ising
model in a non-zero magnetic field $H$ on a random lattice.
This is an example of a theory with a non-trivial matter content.
The action for the spin sector of the theory is chosen as 
\beq{*Ising}
S_I = -\sum_{\{ s_i\}}\big(\sum_{i,j}\beta \delta_{s_i,s_j}
+\sum_i H~s_i\big).
\eeq
The spins are placed in the centers of triangles 
and we use degenerate triangulations
which allow the two vertices to be connected by more than one link and
the links connecting a vertex to itself. This model was solved in 
\cite{boulatov}. For $H > 0$ the model is always in the ordered phase
with the geometric properties of pure gravity ($\gamma_{str}=-1/2$ and
$d_h=4$). For $H=0$ the model has two phases depending on the value of
$\beta$. For $\beta = \beta_c = -\log((2\sqrt{7}-1)/27)$ it undergoes a
third order phase transition between an ordered and a disordered phase.
At the transition the geometric properties change ($\gamma_{str}=-1/3$).
In our experiment we measure the correlation between $\phi_i = (1+s_i)/2$,
(which is 1 for up spin and zero for the down spin). This choice avoids
problems close to $H=0$, where $\la s_i \ra \to 0$ for every $\beta$
on a finite lattice. In our case $\la \phi_i \ra \to 1/2$. The connected
correlator is independent on the additive constant and is (up to a trivial 
factor 1/4) simply the spin-spin correlation function. The distance used
is the triangle distance, i.e. the shortest path on a dual lattice.
We can predict that if the algorithm works, the observed shift $\delta(H)$
should vanish both for large $H$ and for $H \to 0$. For large $H$ the
system becomes completely ordered and the correlation function 
$G_N^{\phi\phi}(r)$
becomes equal to $G_N^{11}(r)$. For $H \to 0$ average spin approaches zero and
$G_N^{1\phi}(r)\approx { 1\over 2}G_N^{11}(r)$.
Below we show results from numerical experiments for system sizes
2k, 4k, 8k and 16k triangles performed at $\beta = \beta_c$ for various
values of the magnetic field $H$. Similar extensive experiments at $H=0$ were
performed by \cite{jmk}, although the definition of distance was different.

\begin{figure}
\begin{center}
%\vspace{5cm}
\includegraphics[width=11cm,bbllx=120,bblly=240,bburx=510,bbury=510]{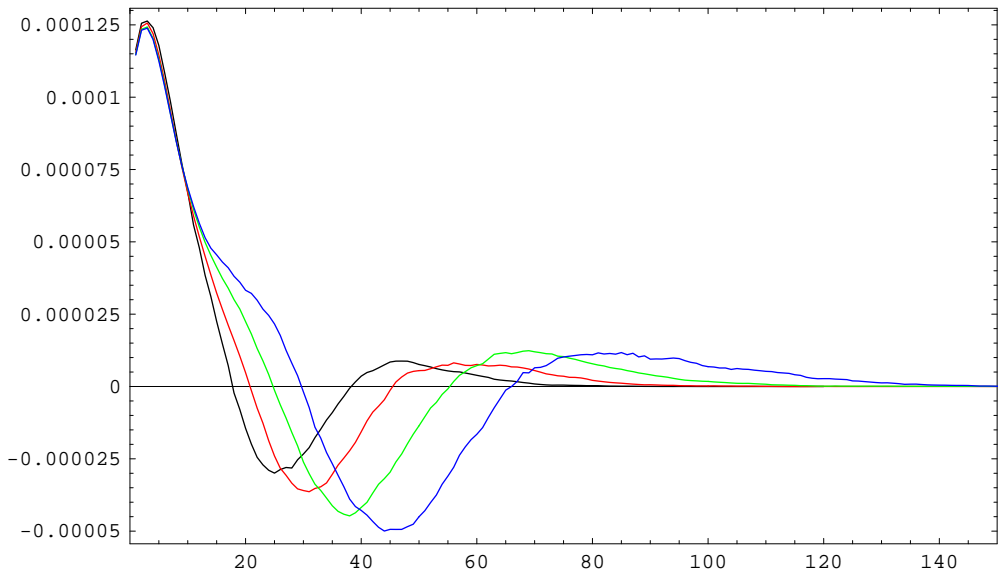}
\end{center}
\caption{\label{all} Spin-spin correlation functions at $H=0.0125$
defined by \rf{*11bis}.}
\end{figure}

On figure \ref{all} we present the connected correlation functions 
$G_N^{\phi\phi}(r)$ versus $r$ for a ``large'' magnetic field $H = 0.0125$
defined in the standard way \rf{*11bis}.
On this and consecutive plots we use colors to distinguish the system 
size: black for 2k, red for 4k, green for 8k and blue for 16k. We see
a large disconnected contribution, which in fact becomes stronger with 
increasing the system size.

On the next plot we show the same functions scaled by a factor 
$N^{3/d_h-1}$ with $d_h=4$ plotted versus the scaling variable 
$x=(r+\delta_0)/N^{1/d_h}$.

\begin{figure}
\begin{center}
\includegraphics[width=11cm,bbllx=120,bblly=240,bburx=510,bbury=510]{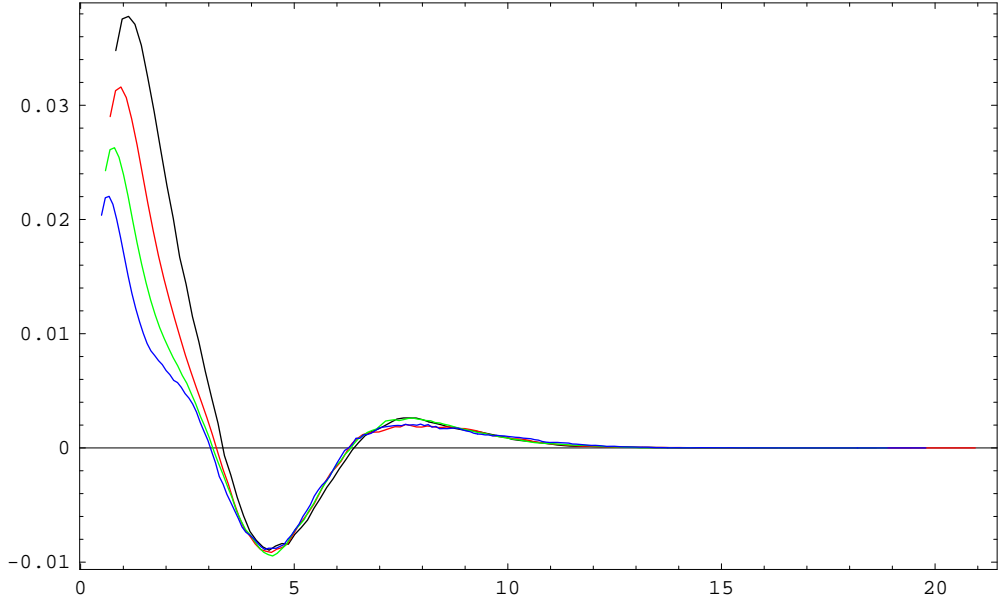}
\end{center}
\caption{\label{all1} Scaled spin-spin correlation functions at $H=0.0125$ }
\end{figure}
\begin{figure}
\begin{center}
\includegraphics[width=11cm,bbllx=120,bblly=240,bburx=510,bbury=510]{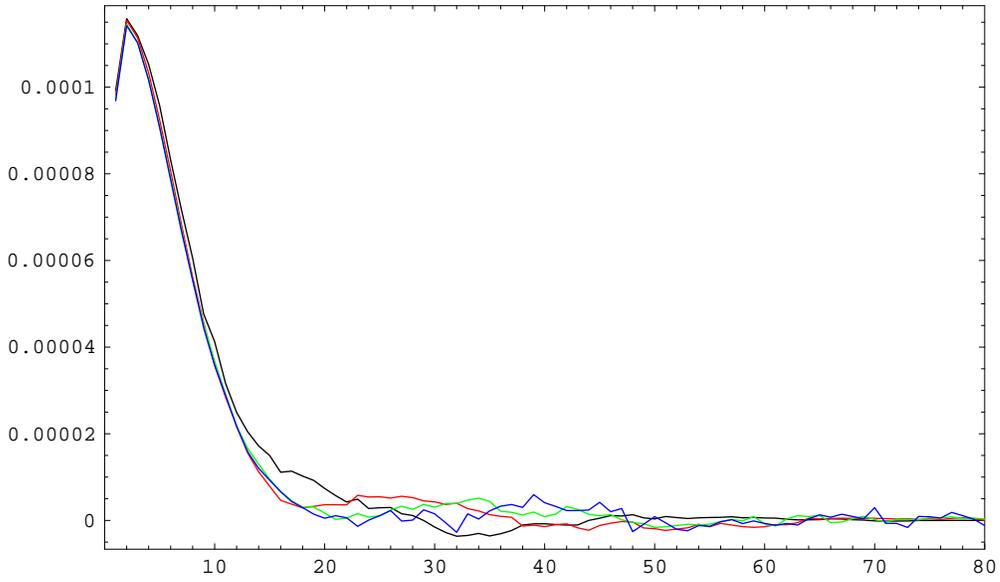}
\end{center}
\caption{\label{non_scale} Spin-spin correlation functions at $H=0.0125$
defined using \rf{*11final}}
\end{figure}

For the sake of presentation we decided
to use the theoretical value $d_h=4$ 
rather than to fit this value. The shift $\delta_0$
was obtained by matching the scaling of the $G_N^{11}(r)$ correlators, again
assuming $d_h=4$.
As we can see the long-range part of the correlator scales exactly as
predicted by \rf{*11again}. 
After reduction by \rf{*11final} the correlators contain
only the non-scaling part. We show it in figure ~\ref{non_scale}. The shift
$\delta(H)= .35\pm .02$ is for this value of $H$ practically independent of the
system size.

\begin{figure}
\begin{center}
\includegraphics[width=11cm,bbllx=120,bblly=240,bburx=510,bbury=510]{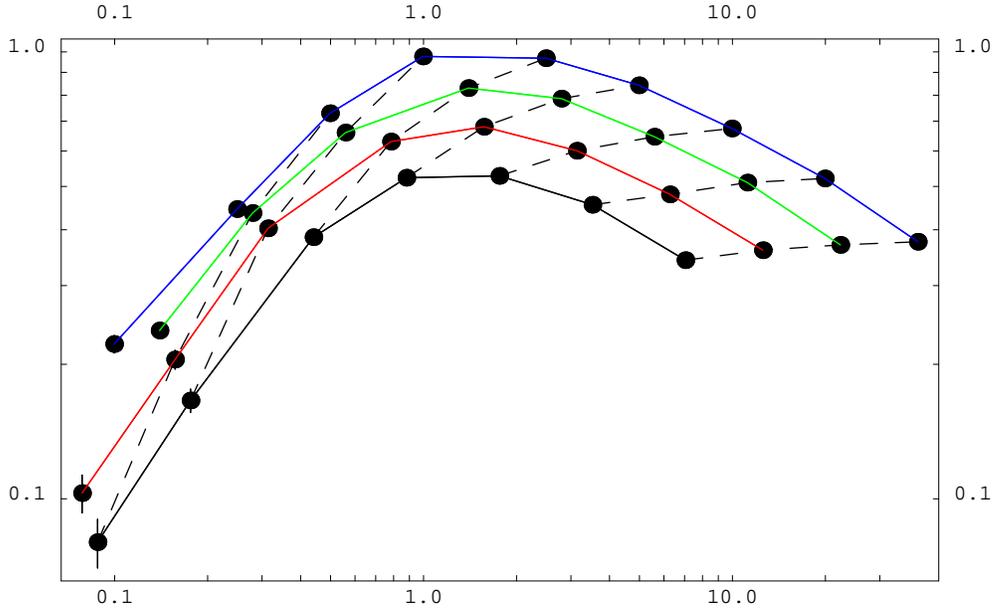}
\end{center}
\caption{\label{delta} Shift $\delta(h)$ vs. $h$.}
\end{figure}

Decreasing the magnetic field $H$ brings us closer to the phase transition.
When we apply the algorithm presented above we discover that for each volume 
the shifts first grow, and then decrease back to zero. The value of the shift
as well as the position of the turn over point start 
to depend on a system volume. From the scaling 
arguments we expect that near the
critical point we should observe a universal behavior of spin-spin
correlations, provided the magnetic field is taken as volume dependent
in the following way\footnote{In flat space we expect 
from standard scaling arguments the singular part of the free energy  
behaves as 
$$ F_{sin}(\th,H) = F_{sin}(1,H/\th^{(\n d + \g_m)/2}),
~~~\th=|\b_b-\b|,$$
where $\g_m$ is the critical exponent for the magnetic field. This formula
is converted into a finite size scaling relation by using that the 
pseudo-critical point for volume $V$ is obtained when the correlation length 
$\xi=\th^{-\n}$ is equal to the linear dimension of the system, i.e.\ to 
$L = V^{1/d}$. Thus
$$ F_{sin}(V,H) = F_{sin}(1, H\,V^{\oh(1+\frac{\g_m}{\n d})}).$$
Thus we expect a universal dependence of $H\,V^{\oh(1+\frac{\g_m}{\n d})}$.
The exponent $\g_m/\n d = 2/3$ after coupling to quantum gravity.}  
\beq{}
H = \frac{h}{N^{5/6}}.
\eeq
On figure \ref{delta} we plot shifts $\delta_N(h)$ for the four system
sizes. We see that the position of the turn-over point corresponds
roughly to $h=1$. Dashed lines join points with the same value of $H$.

Below $h=1$ we observe also a change of $\gamma_{str}$ from the pure 2d 
gravity  value $-1/2$ to the $c=1/2$ value $\gamma_{str}=-1/3$. We measured
this parameter using the standard method of 
measuring the minbu distribution \cite{ajt,ajkt}.

\begin{figure}
\begin{center}
\includegraphics[width=11cm,bbllx=120,bblly=240,bburx=510,bbury=510]{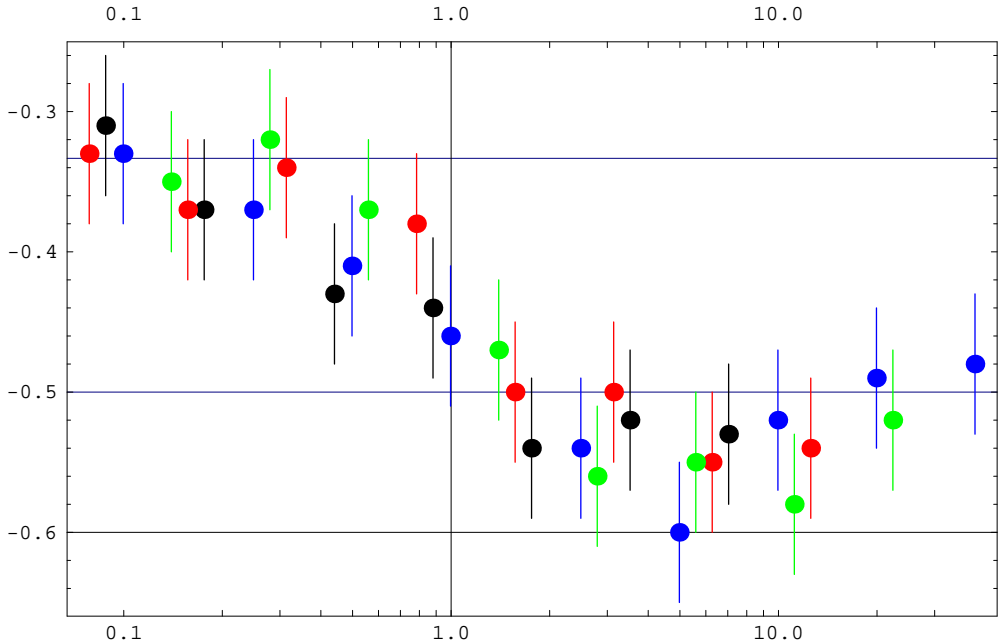}
\end{center}
\caption{\label{gamma} Parameter $\gamma_{str}$ vs. $h$.}
\end{figure}

\begin{figure}
\begin{center}
\includegraphics[width=11cm,bbllx=120,bblly=240,bburx=510,bbury=510]{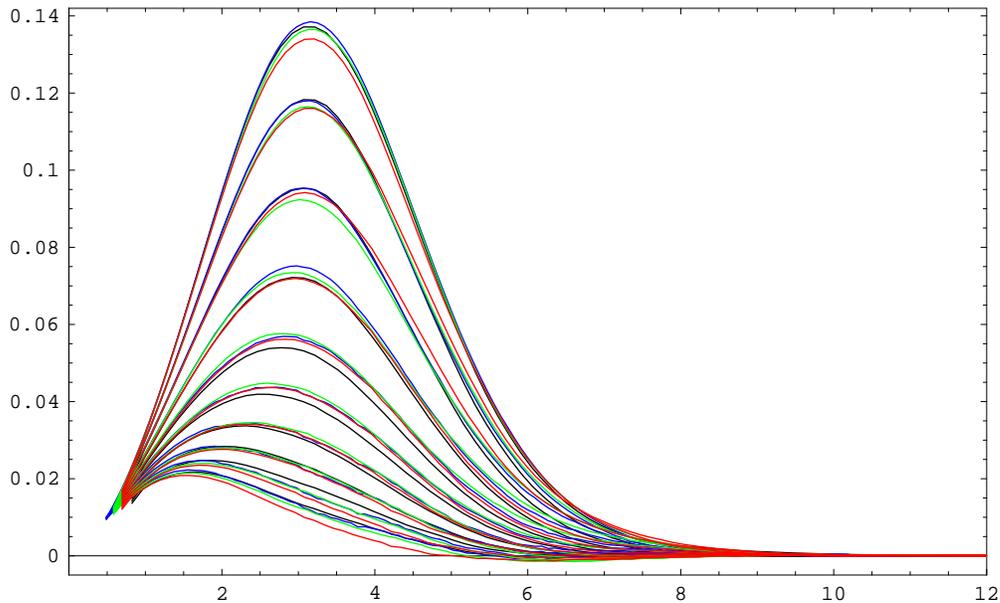}
\end{center}
\caption{\label{hist} Extrapolated spin-spin correlation functions for
(from top to bottom) $h=0.2,~0.4,~\dots,~2.0$}
\end{figure}

Finally figure \ref{hist} represents the correlation functions 
$G_N^{\phi\phi,conn}$ obtained using the improved formula \rf{*11final}, 
scaled by a factor $N^{1/d_h+1/3-1}$ and plotted versus the scaling variable
$x=(r+\delta_0)/N^{1/d_h}$. Values of $d_h=4$ and $\delta_0=4.5$ are the same
as for other plots. Plots correspond from top to bottom
to different values of the scaled
magnetic field $h=0.2,0.4,\dots 2.0$. 
The distributions were obtained in practice by 
interpolating the measured correlation functions at weak fields $H$ to
match the scaled values $h$.
Indeed we observe scaling as predicted.

\section{Discussion}

A reasonable definition of a connected correlator
is given by eq.\ \rf{*4.5a} for a fixed cosmological constant, and
by eq.\ \rf{*11} for a fixed space-time volume. However,
even when no correlations exist for fixed cosmological constant, 
the naive discretized analogue of \rf{*11}, as defined by eq.\ 
\rf{*11bis}, has a non-trivial scaling.
This is made explicit in eq.\ \rf{*11again}.
Basically a dominant disconnected part is still present in 
the definition \rf{*11bis}. As we showed, the refined definition 
\rf{*11final} manages to cancel this dominant disconnected 
part and leaves the genuine connected part of the correlator
as the dominant part of the two-point correlation.
  
\section*{Acknowledgement}
J.A. acknowledges the support 
from MaPhySto, financed by the Danish National Research Foundation.
P.B. was supported by the Alexander von Humboldt
Foundation.
P.B. and J.J. acknowledge the partial support from the KBN grants no.
2P03~B04412 and 2P03~B00814.

\end{document}